\documentclass[10pt]{elsarticle}
\usepackage{graphicx}
\usepackage{caption}
\usepackage{amssymb}
\usepackage{color}

\journal{Journal of Quantitative Spectroscopy \& Radiative Transfer, }

\begin{document}
\begin{frontmatter}
\title{Fractal Signatures in Analogs of \\Interplanetary Dust Particles}

\author{Nisha Katyal$^1$, Varsha Banerjee$^2$ and Sanjay Puri$^1$ \\
$^1$ School of Physical Sciences, Jawaharlal Nehru University,\\ New Delhi -- 110067, India.\\
$^2$ Department of Physics, Indian Institute of Technology, Hauz Khas,\\ New Delhi -- 110016, India.}

\begin{abstract}
Interplanetary dust particles (IDPs) are an important constituent of the earth's stratosphere, interstellar and interplanetary medium, cometary comae and tails, etc. Their physical and optical characteristics are significantly influenced by the morphology of silicate aggregates  which form the core in IDPs. In this paper we reinterpret scattering data from laboratory analogs of cosmic silicate aggregates created by Volten et al. \cite{volten2007}, to extract their morphological features. By evaluating the structure factor, we find that the aggregates are mass fractals with a mass fractal dimension $d_{m} \simeq 1.75$. The same fractal dimension also characterizes clusters obtained from {\it diffusion limited aggregation} (DLA). This suggests that the analogs are formed by an irreversible aggregation of stochastically-transported silicate particles.

\end{abstract}

\begin{keyword}
silicate cores, interplanetary dust particles, structure factor, mass fractals, diffusion limited aggregation. 
\end{keyword}

\end{frontmatter}

\newpage
\section{Introduction}
\label{section1}

Fractal geometries provide a description for many forms in Nature such as coastlines, trees, blood vessels, fluid flow in porous media, burning wavefronts, dielectric breakdown, diffusion-limited-aggregation (DLA) clusters, bacterial colonies, colloidal aggregates, etc. \cite{Mand,Barb,Tam}. They exhibit self-similar and scale-invariant properties at all levels of magnification and are characterized by a non-integer fractal dimension. These features arise because the underlying processes have an element of stochasticity in them. Such processes play an important role in shaping the final morphology, and their origin is distinctive in each physical setting. 

Irregular and rough aggregates have also been observed in the astronomical context. Naturally found cosmic dust aggregates, known as interplanetary dust particles (IDPs), are collected in earth's lower stratosphere. They are formed when dust grains collide in a turbulent circumstellar dust cloud such as the solar nebula, and are an important constituent of the interstellar medium, interplanetary medium, cometary comae and tails, etc. Mass spectroscopy analysis of IDPs have revealed that their primary constituents are (i) silicates of Fe, Mg, Al and Ca, (ii) complex organic molecules of C, H, O and N, (iii) small carbonaceous particles of graphite, coal and amorphous carbon and (iv) ices of CO$_{2}$, H$_{2}$O and NH$_{3}$  \cite{gberg,gberg2,riet2002,tsuch,cuppen}.  Amongst these, there is an exclusive abundance of silicates which aggregate to form particle {\it cores}. They have been  described as {\it fluffy, loosely-structured particles with high porosity}.  The  other constituents contribute to the outer covering or the {\it mantle} and are usually contiguous due to flash heating from solar flares and atmospheric entry \cite{riet1996}. The core, being deep inside retains its morphology. The latter is believed to have a fractal organization characterized by a fractal dimension, but  this belief is not on firm grounds as yet \cite{riet1993,riet2004}.  As the core morphology affects the physical and optical characteristics of IDPs, its understanding has been the focus of several recent works \cite{messenger,min07,henn,rag,botet,nisha}.

Two classes of stochastic fractals are found in nature. The first class is that of {\it surface fractals} whose mass $M$ scales with the radius of gyration $R$ in a Euclidean fashion, i.e., $M \sim R^{d}$, where $d$ is the dimensionality. However, the surface area $S$ increases with the radius as $S \sim R^{d_{s}}$, where $d_{s}$ is the  surface fractal dimension and $d-1\le d_{s} < d$ \cite{adj}. Interfaces generated in fluid flows, burning wavefronts, dielectric breakdown and deposition processes are examples of surface fractals. The second class is that of {\it mass fractals} which obey the scaling relationship, $M \sim R^{d_{m}}$, where $d_{m}$ is the mass fractal dimension and $1 \le d_{m} < d$. Examples of mass fractals are DLA clusters, bacterial colonies and colloidal agregates. Further, in many situations, mass fractals are bounded by surface fractals \cite{Mand,Barb,Tam}.  As a matter of fact, the above mass fractals belong to this class.  

There are many unanswered questions in the context of fluffy cores or silicate aggregates of IDPs. For example, are they mass fractals, bounded by surface fractals? What is their mass and surface fractal dimension? What kind of aggregation mechanisms are responsible for this morphology? What are the consequences of fractal organization on the evolution of clusters? In this paper, we provide answers to some of these questions using the {\it real-space correlation function} $C\left(r\right)$ and the {\it momentum-space structure factor} $S\left(k\right)$. Smooth morphologies are characterized by the {\it Porod law} \cite{op88,pw09}.
The {\it signature} of fractal domains and interfaces is a {\it power-law decay} with non-integer exponents in $C\left(r\right)$ and $S\left(k\right)$. As typical experimental morphologies are smooth on some length scales and fractal on others, the behaviors of  $C\left(r\right)$ vs. $r$ and $S\left(k\right)$ vs. $k$ are characterized by {\it cross-overs} from one form to another. We identify these features in laboratory analogs of cores of IDPs  created by Volten et al. using magnesio-silica grains, by reinterpreting their light-scattering data \cite{volten2007}. We find that these aggregates are mass fractals with a fractal dimension $d_{m} \simeq 1.75$. The same fractal dimension characterizes diffusion limited aggregation (DLA). We therefore infer that aggregation mechanisms of silicate cores in IDPs are stochastic and irreversible as in DLA.   

This paper is organized as follows. In Section~\ref{section2}, we describe the tools for morphology characterization and their usage to obtain mass and surface fractal dimensions. In Section~\ref{section3}, we describe the experimental analogs of silicate cores in IDPs and obtain the structure factor from their light scattering data to extract fractal properties. In Section~\ref{section4}, we present a simulation of the DLA cluster, and the evaluation of its structure factor and the corresponding mass fractal dimension. Finally, we conclude with a summary and discussion of our results in Section~\ref{section5}.

\section{Tools for Morphology Characterization}
\label{section2}

A standard tool to obtain information about sizes and textures of domains and interfaces is the two-point spatial correlation function \cite{pw09}:
\begin{equation}
\label{corfn}
C\left(r \right) =\langle \psi\left(\vec{r_{i}}\right) \psi\left(\vec{r_{j}}\right) \rangle-\langle\psi\left(\vec{r_{i}}\right)\rangle \langle \psi\left(\vec{r_{j}}\right)\rangle,
\end{equation}
where $\psi\left(\vec{r_{i}}\right)$ is an appropriate order parameter and $r = |\vec{r_i} - \vec{r_j}|$.  (We assume the system to be translationally invariant and isotropic.) The angular brackets denote an ensemble average. 

The scattering of a plane wave by a rough morphology can yield useful information about the texture of the domains and interfaces in it. Thus, small-angle scattering experiments (using X-rays, neutrons, etc.) can be used to probe their nature. The intensity of the scattered wave in these experiments yields the momentum-space structure factor, which is the Fourier transform of the correlation function \cite{op88,pw09,sorensen2001,sor1997}:
\begin{equation}
\label{strfac}
S (\vec{k}) = \int \mbox{d}\vec{r} e^{i\vec{k}\cdot\vec{r}}C\left(\vec{r}\right),
\end{equation}
where $\vec{k}$ is the wave-vector of the scattered beam. The properties of $C\left(r \right)$ and $S\left(k\right)$ provide deep insights into the nature of the scattering morphology. 

Consider a domain of size $\xi$ formed by spherical particles of size $a$, as depicted schematically in Fig. 1(a). The typical interfacial width $w$, is also indicated. This prototypical morphology  could represent a colloidal aggregate, soot particles, a DLA cluster, etc. The correlation function for such a  morphology can be  approximated by
\begin{eqnarray}
\label{corrf}
1 - C\left(r \right) = \bar{C}\left(r\right)  \simeq \left\{\begin{array}{ll}
A r^{\alpha}, & \quad w \ll r \ll \xi,\\
B r^{\beta},   & \quad r \ll w \ll a,\\
C r^{\gamma} & \quad  r \ll a.
\end{array} 
\right.
\end{eqnarray}
The first term conveys information about the domain texture probed by length scales $w \ll r \ll \xi$. If the domain has no internal structure, $\alpha$ = 1 signifying Porod decay \cite{op88,pw09}. For a fractal domain, on the other hand,  $\alpha$ = $d_{m} - d$ where $d_{m}$ is the mass fractal dimension \cite{sorensen2001,sor1997}. The second term conveys information about the properties of the interface, probed by lengths $a \ll r \ll w$. For fractal interfaces, $0 \le \beta < 1$, and $\beta$ is related to the fractal dimension as $d_{s} = d - \beta$ \cite{dmp}. The third term is significant only if the building blocks are particles of diameter $a$. In that case, $\gamma = 1$ for $r \lesssim a$, yielding a Porod regime at a microscopic length scale.

In Fourier space, Eq.~(\ref{corrf}) translates into the following power-law behavior of the structure factor:
\begin{eqnarray}
\label{sf}
S\left(k\right)  \simeq \left\{\begin{array}{ll}
\tilde{A} k^{-(d+\alpha)}, & \quad \xi^{-1} \ll k \ll w^{-1},\\
\tilde{B} k^{-(d+\beta)},   & \quad w^{-1} \ll k \ll a^{-1},\\
\tilde{C} r^{-(d+\gamma)} & \quad a^{-1} \ll k.
\end{array} 
\right.
\end{eqnarray}
The Porod decay of the form $k^{-\left(d+1\right)}$ in the scattered intensity is typical of smooth domains or sharp interfaces \cite{op88,pw09}. A deviation from this behavior to $S(k) \sim k^{-(d \pm \theta)}$ is indicative of a fractal structure in the domains or interfaces. When physical structures have multiple length-scales, one or more terms in Eqs.~(\ref{corrf}) and (\ref{sf}) may contribute. Their presence is characterized by cusps in the correlation function, and corresponding power-laws in the structure factor \cite{sbp}.

We illustrate the power laws and cross-overs discussed above in the context of the 2-$d$ morphology depicted in Fig. 1(a). It should be noted that both the domain and the interfacial boundary in this schematic are rough, self-similar fractals. The structure factor $S\left(k\right)$ vs. $k$ for this morphology obtained from the Fourier transform of the spherically-averaged correlation function $C\left(r\right)$ vs. $r$ is plotted in Fig. 1(b) on a log-log scale. This function exhibits two distinct regimes over large and small values of $k$ as seen from the best fit lines: {\it power-law decay} with $S\left(k\right)\sim k^{-1.71}$ for $\xi^{-1} \ll k \ll w^{-1}$ and a  {\it Porod decay} with $S\left(k\right)\sim k^{-3}$ for $a^{-1} \ll k$. With reference to Eqs.~(\ref{corrf}) and (\ref{sf}), the power law decay signifies a fractal domain morphology with a mass fractal dimension $d_{m} \approx 1.71$ while the Porod  decay is due to the smooth surface of the particles. The structure factor corresponding to wave vectors in the interval $w^{-1} \ll k \ll a^{-1}$ is due to scattering from the rough fractal  interfaces. However it difficult to identify the corresponding power law with precision due to cross-overs effects from the adjoining mass fractal and Porod regimes.

\section{Analysis of Silicate Cores}
\label{section3}

We now investigate the morphological characteristics of silicate cores using the correlation function and the structure factor.  As real samples are scarce, it has been customary to create them  in the laboratory using a condensation flow apparatus followed by flash heating to mimic the environment required for the formation of cosmic silicates and circumstellar dust. A significant contribution in this context is the work of Volten et al. \cite{volten2007}. They created a variety of magnesio-silica samples with (relative) concentrations typical of silicate cores in IDPs \cite{frim}. The mixed grains in these samples formed interconnected, tangled chains ranging from open structures to dense structures, thereby yielding samples of varied porosities. We calculate $S\left(k\right)$ for two such analogs:  Sample 1 has  an equal proportion of Mg and Si; and Sample 2 has Mg and Si in the ratio 1.4:1. These samples have a porosity of $\sim 40\%$ \cite{volten2007,frim}.

\begin{figure}
\centering
\begin{minipage}{.5\textwidth}
  \centering
\vspace*{0.6cm}
  \includegraphics[width=.96\linewidth]{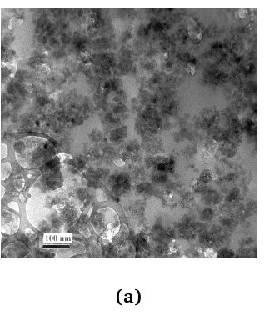}
\end{minipage}%
\begin{minipage}{.5\textwidth}
  \centering
\vspace*{0.3cm}
  \includegraphics[width=1.1\linewidth]{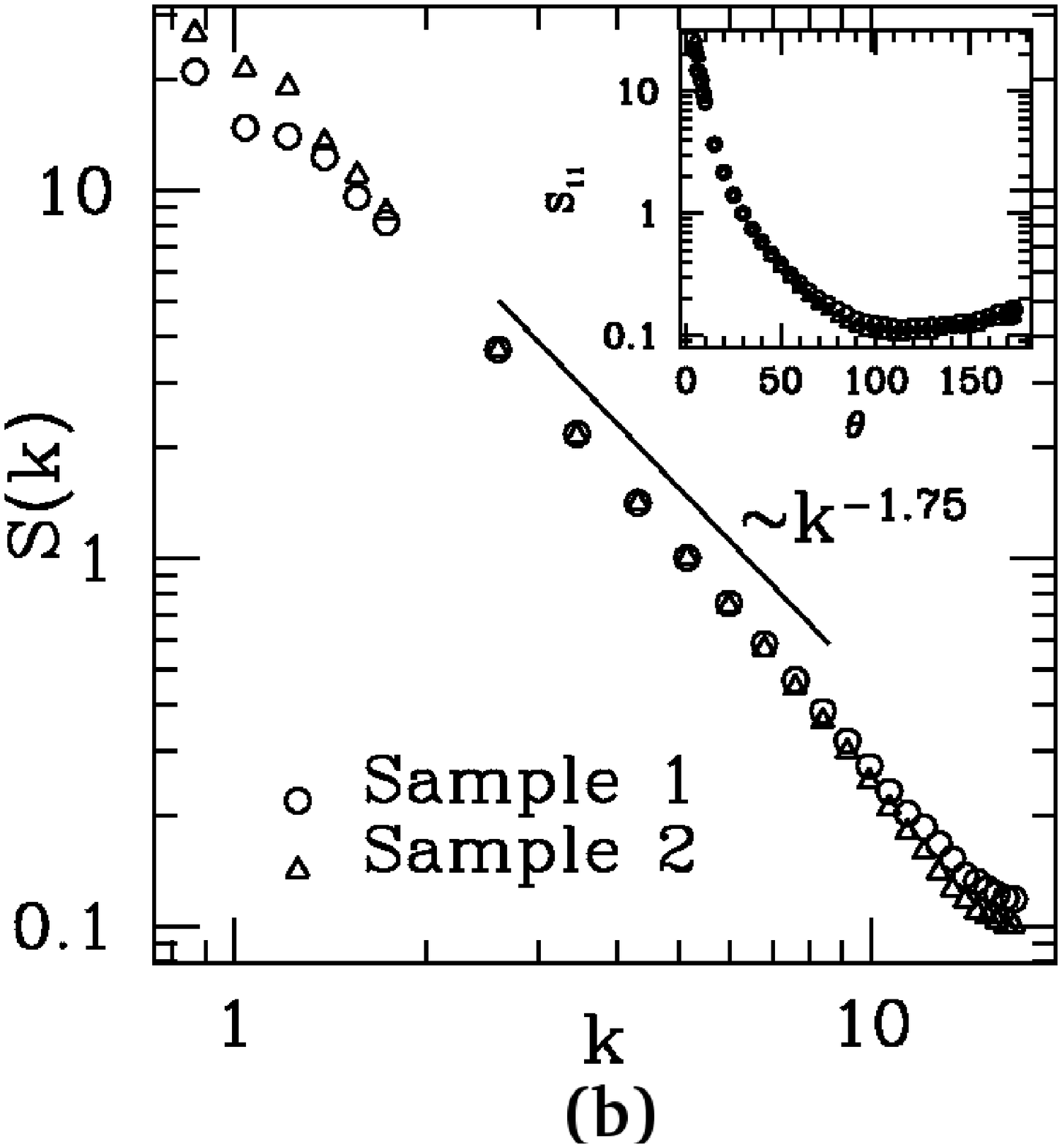}
\end{minipage}
\caption{(a) TEM image of a section sliced through a fluffy MgSiO
particle consisting of several small aggregates of magnesio-silica
grains forming a contiguous, yet porous structure \cite{volten2007}. (b) Scattering intensity $S(k)$
as a function of scattering wave-vector $k$ for Samples 1 and 2 on a
log-log scale. A line of slope $-1.75$ fits well to the intermediate-$k$
data. The original experimental data from Ref.~\cite{volten2007},
showing the variation of scattering phase function $S_{11}$ as a
function of scattering angle $\theta$, is provided in the inset.}
\label{f2}
\end{figure}

Fig.~\ref{f2}(a) reproduces a prototypical TEM image of an ultra-thin section sliced through MgSiO particles prepared by Volten et al. (The image is reproduced from \cite{volten2007} with permission from the authors.) They are organized in the form of small fluffy aggregates organized in a contiguous but porous morphology \cite{volten2007} Volten et al.  then obtained light-scattering data for the samples using the Amsterdam light scattering database. The light-scattering properties were measured at a wavelength of 632.8 nm with the range of scattering angles from $5^{\circ}$ to $174^{\circ}$, in  steps of $1^{\circ}$. These measurements yielded the scattering matrix elements as a function of scattering angle $\theta$. The inset of Fig.~\ref{f2}(b) plots  scattering phase function $S_{11}$ vs. $\theta$. We convert this data in terms of the magnitude of the scattering wave-vector by the transformation $k=4\pi/\lambda \ \mbox{sin}(\theta/2)$. The transformed data sets are presented in Fig.~\ref{f2}(b) on a log-log scale. The intermediate-$k$ region is linear on this plot, implying a power-law dependence between the scattering intensity $S(k)$ and the scattering wave-vector $k$. The best-fit line (shown alongside) has a slope of $-1.75$. From Eqs.~(\ref{corrf})-(\ref{sf}) and the accompanying discussion, it is clear that the fluffy aggregates of Samples 1 and 2 are mass fractals with a mass fractal dimension $d_{m} \simeq 1.75$.    

\section{Diffusion Limited Aggregation}
\label{section4}

A relevant question now is: What kind of mass-transport mechanisms lead to fluffy aggregates with $d_{m} \simeq 1.75$? To answer this question, we create aggregates of particles using the DLA model. We performed this simulation on a cubic lattice adopting the algorithm introduced by Meakin \cite{meakin}: (i) A particle is placed at the origin or the center of the cube. (ii) A new particle is released at a distance $R$ from the center and performs a random walk. (iii) On encountering  an occupied neighboring site, it adheres irreversibly to it. Steps (ii) and (iii) are repeated several times to obtain a DLA cluster.

To mimic the morphology (of several small aggregates) observed in the TEM micrograph of  Fig.~\ref{f2}(a), we simulate a DLA cluster ($d=3$) using multiple seeds. Each seed initiates a sub-cluster using the above procedure. We allow the sub-clusters to grow till they form a contiguous, yet delicately branched, self-similar structure. Fig.~3(a) depicts such a prototypical multi-seed cluster built from $\sim 10^{4}$ particles. We also show a slice of this morphology ($d = 2$) in Fig.~3(b). It contains two initial seeds, marked in red. Notice here the similarity of this slice with the open contiguous structure in the TEM image of Fig.~\ref{f2}(a).  Next, we quantify the morphology of Fg.~3(a) by evaluating the spherically-averaged structure factor $S(k)$, which is shown in Fig.~3(c) on a log-log scale. The power-law behavior at intermediate values of $k$ fits best to a line with slope $-1.75$. With reference to Eq.~(\ref{sf}) and the discussion thereafter, DLA clusters are mass fractals with $d_{m} \simeq 1.75$. In view of these observations, we infer that the aggregates obtained in the experiments of Volten et al. with $d_{m} \simeq 1.75$ are due to irreversible aggregation of stochastically transported silicate particles.


\section{Conclusion}
\label{section5}

Interstellar dust particles (IDPs) found in the earth's stratosphere are an important constituent of cosmic matter. These comprise of loosely structured silicate cores or aggregates ensconced in a mantle of organic and carbonaceous compounds \cite{gberg,gberg2,riet2002,tsuch,cuppen}. The organization and optical characteristics of the IDPs are greatly influenced by the morphology of the core. In this paper, we have re-interpreted scattering data from laboratory analogs of silicate cores in IDPs created by Volten et al. \cite{volten2007} using the correlation function $C(r)$ and the structure factor $S(k)$. This analysis has provided us a means to quantify characteristics such as the size and texture of these aggregates. The presence of fractal architecture is characterized by power laws with non-integer exponents in the structure factor. We found that the silicate aggregates are mass fractals with a fractal dimension $d_{m} \simeq 1.75$. This value of $d_m$ is the same as the fractal dimension of aggregates obtained in a diffusion limited aggregation model. We therefore conclude that the aggregates are formed by an irreversible aggregation of stochastically transported silicate particles. We have also studied the effect of density of particles on the fractal dimension. Our observation is that $d_{m}$ approaches the Euclidian dimension ($d = 3$) with increasing density.  Further, we wish to emphasize that $C(r)$ and $S(k)$ contain information averaged over {\it all} domains and interfaces in contrast to the conventionally used (local) box-counting procedures. Our estimates of $d_{m}$ are therefore very accurate.   

For confirmation of our results and further insights, we require data from real cosmic dust and cometary particles. We understand that it is difficult to obtain light scattering data from stellar objects. Alternatively, information providing depth profiles of these assemblies could also be used to evaluate the $C(r)$ and $S(k)$. As discussed above, they are excellent tools for morphology characterization especially due to their direct experimental relevance. More generally, micro-scale phenomena are characterized by mass-dependent diffusion, i.e., the diffusion rate $D(m) \sim m^{-\alpha}$, where $m$ is the mass or number of particles in the cluster and $\alpha$ is a system-specific parameter \cite{sbp}. The fractal characteristics of aggregates are greatly influenced by $\alpha$. We are presently investigating them to quantify this influence. We hope that such analyses will enhance our understanding of diffusion mechanisms in mega-scale systems found in the cosmic environment. 

\section*{Acknowledgements}\label{sec:acknow}
The authors would like to thank the anonymous refeeres for their constructive comments that helped to improve the quality of the paper. VB would like to acknowledge the support of DST Grant No. SR/S2/CMP-002/2010.

\newpage

\newpage 
\begin{figure}
\centering
\begin{minipage}{.5\textwidth}
  \centering
\vspace*{-0.15cm}
\hspace*{-1cm}
  \includegraphics[width=1.0\linewidth]{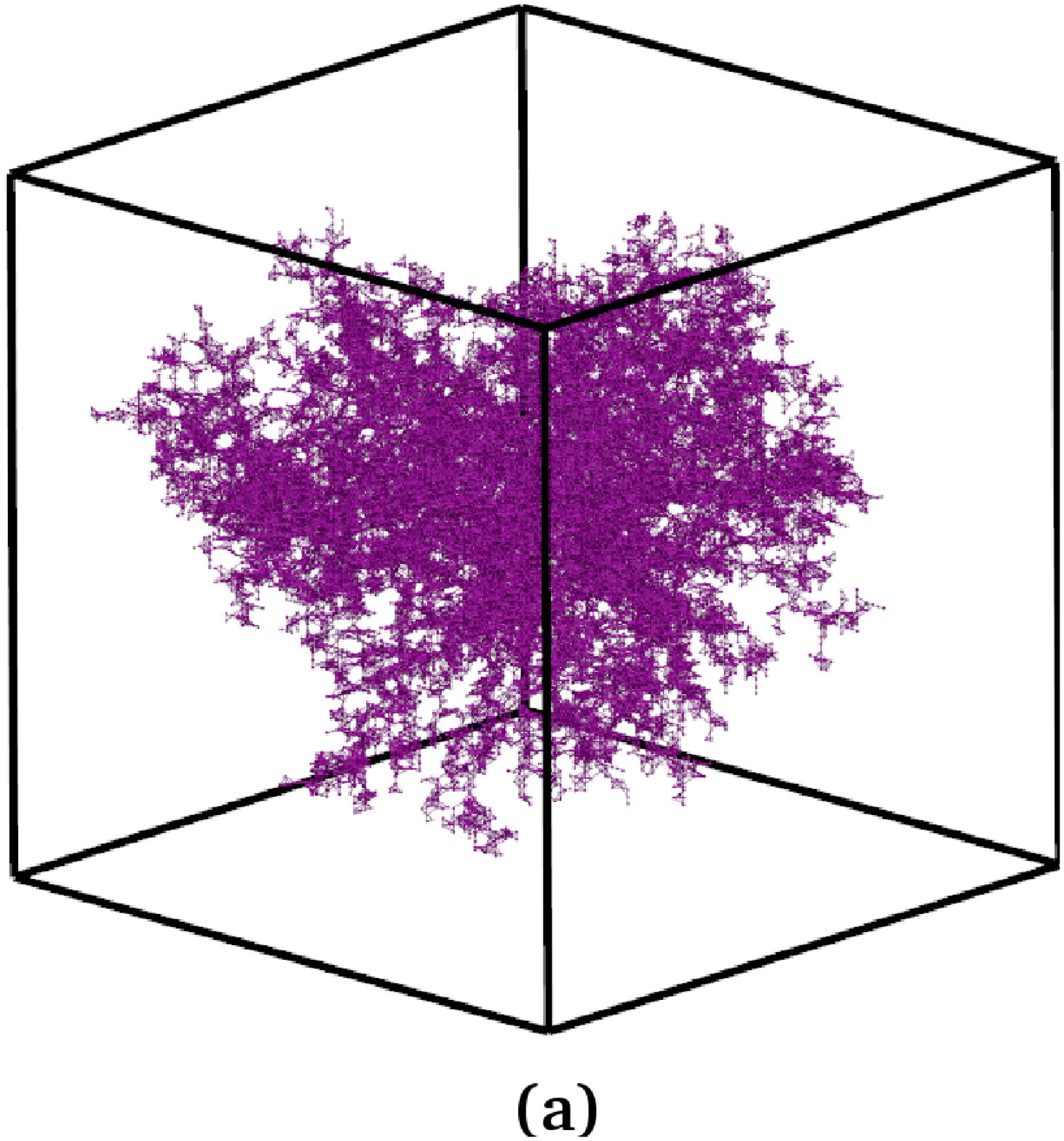}
\end{minipage}%
\begin{minipage}{.5\textwidth}
  \centering
\vspace*{-0.15cm}
\hspace*{-0.6cm}
 \includegraphics[width=0.79\linewidth]{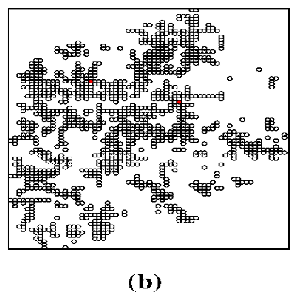}

\end{minipage}
\begin{minipage}{.5\textwidth}
  \centering
\vspace*{0.3cm}
\hspace*{-1.1cm}
  \includegraphics[width=1.20\linewidth]{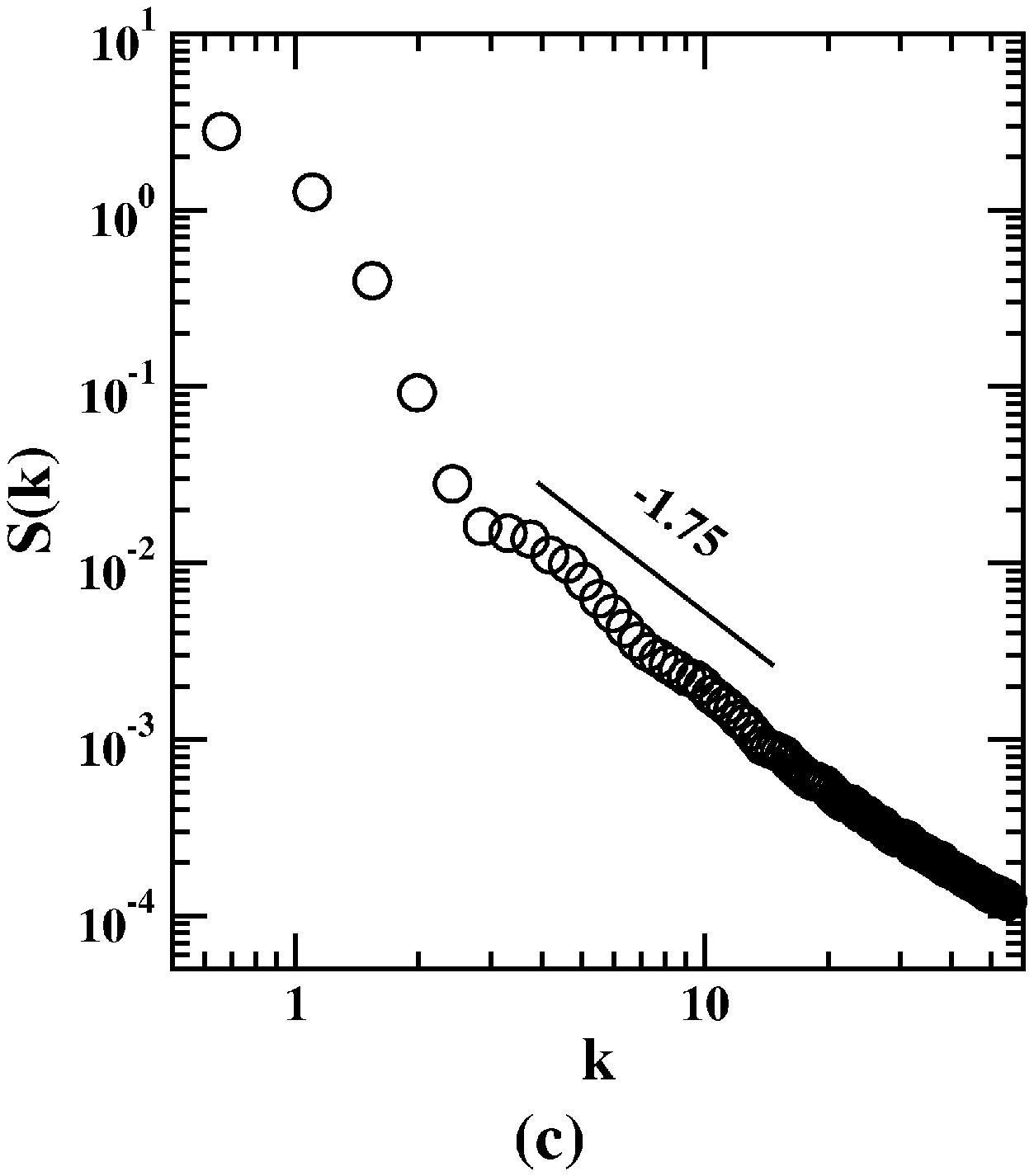}
\end{minipage}
\caption{(a) Computer generated DLA cluster ($d = 3$) with multiple seeds obtained from $\sim 10^{4}$ particles. (b) A slice ($d = 2$  of the DLA cluster depicting two initial seeds (red) and the delicately branched, self-similar contiguous growth.}   (c) Spherically averaged structure factor, $S(k)$ vs. $k$, on a log-log scale for the DLA cluster in (a). The power-law behavior in the intermediate-$k$ regime yields the mass fractal dimension $d_m \simeq 1.75$.
\label{f3}
\end{figure}


\begin{thebibliography}{00}

\bibitem{volten2007} Volten, H., Munoz, O., Hovenier, J.W., Rietmeijer, F.J.M., Nuth, J.A., Waters, L.B.F.M., et al, 2007, Astron Astrophys, 470, 377.

\bibitem{Mand} Mandelbrot, B. B., \textit{The Fractal Geometry of Nature} (W. H. Freeman, 1982).

\bibitem{Barb} Barabasi, A. L. and Stanley, H. E., \textit{Fractal Concepts in Surface Growth} (Cambridge University, 1995).    
\bibitem{Tam} Vicsek, T., \textit{Fractal Growth Phenomena} (World Scientific, 1992).      

\bibitem{gberg} Greenberg, J. M., 1989, From interstellar dust to comet dust and inteplanetary particles, in Highlights of Astronomy, Vol. 8, 241–250.

\bibitem{gberg2} Greenberg, J. M., 1998, Earth, Moon and Planets, Vol. 82-83, Issue 0, pp 313-324. 

\bibitem{riet2002} Rietmeijer, F. J. M., 2002, Chem. Erde, 62, 1.

\bibitem{tsuch} Tsuchiyama, A., Uesugi, K., Nakano, T., Okazaki, T., Nakamura. K, Nakamura. T., Noguchi, T. and Yano, H., 2006, Annual Lunar and Planetary Science Conference XXXVII, Texas, abstract no. 2001.

\bibitem{cuppen} Cuppen, H. M., and E. Herbst, 2007, Simulation of the Formation and Morphology of Ice Mantles on Interstellar Grains, ApJ, 668, 294.

\bibitem{riet1996} Rietmeijer, F. J. M., 1996, Meteoritics Planet Sci., 31, 237.

\bibitem{riet1993} Rietmeijer, F. J. M., 1993, Earth Planet. Sci. Lett., 117, 609.

\bibitem{riet2004} Rietmeijer, F. J. M. and Nuth III, J. A., 2004, ASSL Vol. 311: The New Rosetta Targets. Observations, Simulations and Instrument Performances, ed. L. Colangeli, E.M. Epifani, and P. Palumbo (Astrophys. Space Sci. Library, Kluwer Academic Publishers), 97-110.

\bibitem{messenger} Messenger, S., Keller, L. P., Stadermann, F. J., Walker, R. M., Zinner, E., 2003, Science, 300, 105.


\bibitem{min07} Min, M., Waters, L. B. F. M., Koter, A. de., Hovenier, J. W., Keller, L. P., Markwich-Kemper, F., 2007, A \& A, 486, 779.

\bibitem{henn} Henning, T., 2010, Annu. Rev. Astron. Astrophys., 48, 21-46.

\bibitem{rag} Vaidya, D. B. and Gupta, R., 2011, A\&A, 528, A57.

\bibitem{botet} Botet, R. and Rakesh R., 2013, Earth Planets Space, Vol. 65 (No. 10), pp. 1133.

\bibitem{nisha} Katyal, N., Gupta, R. and Vaidya, D. B., 2013, PASP, Vol. 125, 1443.

\bibitem{adj} Hurd, A. J., Schaefer, D. W. and Martinn, J. E., 1987, Phys. Rev. A, 35, 2361.

\bibitem{op88} Porod, G., in {\it Small-Angle X-Ray Scattering}, edited by O. Glatter and O. Kratky (Academic Press, New York, 1982); Oono, Y. and Puri, S., Mod. Phys. Lett. B 2, 861 (1988).

\bibitem{pw09} {\it Kinetics of Phase Transitions}, edited by S. Puri and V.K. Wadhawan, Taylor and Francis, Boca Raton (2009).

\bibitem{sorensen2001} Sorensen, C. M., 2001, Aerosol Sci. Tech., 35, 648.

\bibitem{sor1997} Oh, C. and Sorensen, C. M., 1997, Phys. Rev. E, 193, 17.

\bibitem{dmp} Mildner, D. R. R. and Hall, P. L., 1986, J. Phys. D: Appl. Phys., 19, 1535.

\bibitem{sbp} Shrivastav, G.P., Banerjee, V. and Puri, S. 2010, Eur. Phys. J. B, 78, 217.


\bibitem{frim} Rietmeijer, F. J. M. (1998) Interplanetary Dust Particles. In Planetary Materials, Reviews in Mineralogy, vol. 36 (J.J. Papike, ed.), 2-1 – 2-95, Mineralogical Society of America, Chantilly, Virginia.

\bibitem{meakin} Meakin, P., 1983, Phys. Rev. A, 27, 1495.

\newpage
\begin{figure}
\vspace*{-0.5cm}
\hspace*{-0.5cm}
\includegraphics[width=.96\linewidth,angle=270]{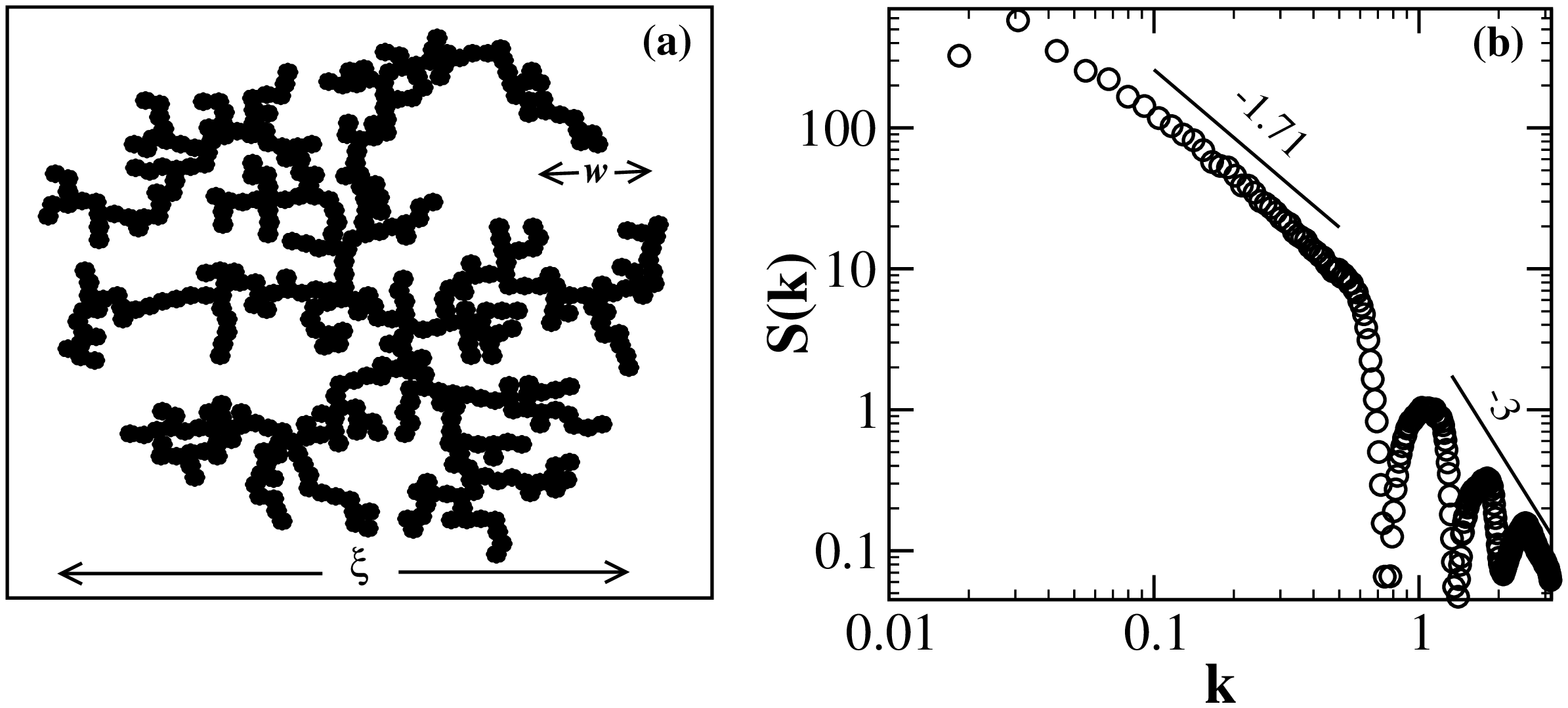}
\caption{(a) A typical morphology of a domain of size $\xi$ formed by spherical particles of size $a$ is depicted. (b) Log-log plot of structure factor of the morphology as shown in (a). A power-law decay with a slope of -1.71 signifies a fractal morphology whereas a slope of -4 signifies a Porod law decay due to smooth morphology of the overall domain at that particular length scale.}
\end{figure}

\end{thebibliography}
\end{document}